 \documentclass[final,3p,times]{elsarticle}

\usepackage{amssymb,amsmath,amsthm, amsfonts, amstext}
\usepackage{footnote}
\usepackage{rotating}
\usepackage{hhline}
\usepackage[normalem]{ulem}
\usepackage{graphicx}
\usepackage{chngcntr}
\usepackage{color}
\usepackage{lineno}

\newcommand{\R}{\mathbb{R}}

\def \v {\mathrm{v}}
\def \w {\mathrm{w}}
\counterwithout{equation}{section}

\journal{}

\begin{document}

\begin{frontmatter}

\title{Inventory growth cycles with debt-financed investment}

\author[a1]{Matheus R. Grasselli}
\author[a2]{Adrien Nguyen-Huu}

\address[a1]{McMaster University, Hamilton, Canada}
\address[a2]{Universit\'e de Montpellier, Montpellier, France}

\begin{abstract}
We propose a continuous-time stock-flow consistent model for inventory dynamics in an economy with firms, banks, and households. On the supply side, firms decide on production based on adaptive expectations for sales demand and a desired level of inventories. On the demand side, investment is determined as a function of utilization and profitability and can be financed by debt, whereas consumption is independently determined as a function of income and wealth. Prices adjust sluggishly to both changes in labour costs and inventory. Disequilibrium between expected sales and demand is absorbed by unplanned changes in inventory. This results in a five-dimensional dynamical system for wage share, employment rate, private debt ratio, expected sales, and capacity utilization. We analyze two limiting cases: the long-run dynamics provides a version of the Keen model with effective demand and varying inventories, whereas the short-run dynamics gives rise to behaviour that we interpret as Kitchin cycles.   
\end{abstract}


\end{frontmatter}



\section{Introduction}
\label{sec:introduction}

Inventory fluctuations have been known for a long time to be a major component of the business cycle \cite{Abramovitz1950}. According 
to \cite{Blinder1981}, even though investment in inventory accounts for a very small fraction of output (about 1 percent in the U.S.), changes in 
inventory investment account for a disproportionately large fraction of changes in output over the cycle (about 60 percent on 
average for seven postwar recessions in the U.S.). Nevertheless, inventory dynamics has received relatively little attention in 
the theoretical literature. A review of earlier models is provided in \cite{BlinderMaccini1991}, where it is observed that, whereas ``the prevailing 
micro theory viewed inventories as a {\em stabilizing} factor'', the data shows that output is more volatile than final sales (namely output less 
inventory investment), suggesting a destabilizing role for inventories in macroeconomics. The landscape has not changed 
significantly since then, with a few recent papers focussed on incorporating inventories in fully micro founded general equilibrium models \cite{WangWenXu2014, Wen2011}. As remarked in these papers, explaining inventories in a frictionless general equilibrium model is as challenging as explaining money, forcing 
this type of analysis to rely on frictions, such as delivery costs and stockout-avoidance motives, akin to the attempts to incorporate a financial sector into DSGE models. 
In this paper, we follow an alternative approach based on disequilibrium models where sluggish adjustment, adaptive expectations, and sectoral averages replace market 
clearing, rational expectations, and representative agents \cite{GodleyLavoie2007}. 

Our starting point is the growth cycles model proposed in \cite{Franke1996} along the lines originally formulated in \cite{Metzler1941}: investment in inventory 
adjusts to a desired inventory-to-expected-sales ratio, whereas expected sales themselves adapt taking into account fluctuating demand. As shown in \cite{Franke1996}, the interplay between the long-run growth trend and short-run adjustment of inventory stock and expected sales determine the stability of the model. Whereas sufficiently sluggish adjustments promote stability, the model exhibits dynamic instability if the adjustment speeds exceed certain thresholds. Moreover, a flexible inventory adjustment speed can lead to persistent cyclical behaviour. The model in \cite{Franke1996} is described by means of a two-dimensional dynamical system with normalized expected sales and inventory levels (or equivalently, capacity utilization) as state variables and therefore necessarily neglects several other macroeconomic dynamic feedback channels. In particular, the model takes the wage share of the economy as constant, so that endogenous cycles arising from distributional conflict {\em \`a la} Goodwin \cite{Goodwin1967} are not considered. Moreover, in the absence of an explicit financial sector, there is no role in the model for the kind of Minskyan instability \cite{Minsky1982} arising from debt-financed investment.

In this paper, we present in Section \ref{sec:general} a unified model with both inventory and labour market dynamics, as well as allowing for financial considerations to play a role in investment decision through profits net of debt servicing. The resulting dynamics leads to the five-dimensional system derived in Section \ref{sec:main}, with the traditional wage share and employment rate variables of the Goodwin model \cite{Goodwin1967} augmented by the debt ratio of firms as in the Keen model \cite{Keen1995}, in addition to the expected sales and capacity utilization variables of the Franke model \cite{Franke1996} mentioned above. Global analysis of such high-dimensional nonlinear system is beyond the scope of current techniques, and even local analysis of the interior equilibrium proves to be laborious and not very illuminating. In the remainder of the paper we opt instead to investigate two well defined limiting cases.  

In Section \ref{sec:effective demand in the keen model} we consider the type of long-run dynamics that arises when firms have no planned investment in inventory and make no adjustments for short term fluctuations in inventory and expected sales. The resulting model is thus four-dimensional, as the ratio of expected sales 
to output is now constant. Further simplification is achieved by specifying the long-run growth rate of expected sales. When this is chosen to be a constant as in 
\cite{Franke1996}, we find that the labour market part of the model only achieves equilibrium for a particular initial condition for employment. Alternatively, choosing the 
long-run growth rate of expected sales to be the same as the instantaneous growth rate of capital leads to a three-dimensional model for wage share, employment rate, and debt ratio, with a constant capital utilization. The non-monetary version of this model is very similar to the original Keen model, but now with a non-trivial effective demand and fluctuating inventories. As in the Keen model, an equilibrium with infinite debt ratio is also possible but highly problematic in this model, because it leads to infinitely negative inventory levels. When the model is cast in nominal terms, we find that the equilibrium with explosive debt is no longer possible, essentially because the positive wealth effect in the consumption function raises demand and consequently the inflation rate. On the other hand, in addition to a deflationary state first observed in \cite{GrasselliNguyenHuu2015}, a new type of debt crisis corresponding to vanishing wage share and employment rate but with a {\em finite} debt ratio arises as a possible stable equilibrium.

In Section \ref{sec:short-run} we turn our attention to the opposite limiting case, namely a no-growth regime where the only drivers of expected sales and inventory investment are short-run fluctuations in demand. Further ignoring the wealth effect in the consumption function allows us to focus exclusively on a reduced two-dimensional 
describing the relationship between demand and expected sales. This fascinating system undergoes a bifurcation from a locally stable equilibrium in which demand  equals expected sales to an unstable limit cycle. We find that the higher the adjustment speeds of inventory and expected sales, the harder it is to achieve stability. On the other hand, stability is enhanced when prices react faster to mismatch between demand and expected sales. The interplay between supply, demand, and prices is 
even more involved when we consider an alternative model in which prices adjust indirectly through the mismatch between actual and desired inventory levels, rather then directly through changes in inventory. In this case, the increased information lags give rise to stable limit cycles that strongly resemble the Kitchin cycles first reported in \cite{Kitchin1923}.

\section{The General Model}
\label{sec:general}

\subsection{Accounting structure}
\label{sec:general model}

We consider a three-sector closed economy consisting of firms, banks, and households.
The firm sector produces one homogeneous good used both for consumption and investment. 

\paragraph{Capital and utilization}
Denote the total stock of capital in the economy in real terms by $K$ and assume that it determines potential 
output $Y_{p}$ according to the relationship 
\begin{equation}
\label{eq:Y_max}
Y_{p}=\frac{K}{\nu},
\end{equation}
where $\nu$ is a constant capital-to-output ratio. The actual output produced by firms $Y$ is assumed to consist of 
expected sales $Y_e$ plus planned inventory changes $I_p$, that is,
\begin{equation}
\label{eq:supply}
Y=Y_e+I_p,
\end{equation}
and in turn determines capacity utilization as
\begin{equation}
\label{eq:utilization}
u=\frac{Y}{Y_p}.
\end{equation}
Finally, capital is assumed to change according to 
\begin{equation}
\label{eq:capital}
\dot{K} = I_k - \delta(u) K,
\end{equation}
where $I_k$ denotes capital investment in real terms and $\delta(u)$ is a depreciation rate expressed 
as a function of capital utilization $u$.

\paragraph{Effective demand and inventories}
Denote total real consumption by banks and households by $C$, which together with capital investment $I_k$ determine total sales demand

\begin{equation}
\label{eq:demand}
Y_d=C+I_k,
\end{equation}
also in real terms. The difference between output and demand determines actual changes in the level of inventory held by firms. 
In other words, 

\begin{equation}
\label{eq:inventory}
\dot{V} = I_p+I_u = Y-Y_d,
\end{equation}
where $V$ denotes the stock of inventories and $\dot{V}$ denotes investment in inventory, 
which consists of both planned and unplanned changes in inventory, denoted by $I_p$ and $I_u$ respectively. 
Substituting \eqref{eq:supply} into \eqref{eq:inventory}, we see that unplanned changes in inventory are given by 

\begin{equation}
\label{eq:unplanned}
I_u=\dot{V}-I_p=(Y-Y_d)-(Y-Y_e)=Y_e-Y_d,
\end{equation}
and therefore accommodate any surprises in actual sales compared to expected sales. 
Finally, total real investment in the economy is given by 
$
I=Y-C=Y-Y_d+I_k=I_p+I_u+I_k$,
that is, it consists of changes in inventory, both planned and unplanned, plus capital investment.

\paragraph{Labor cost and employment}
Let the nominal wage bill be denoted by $W$, the total workforce by $N$ and the number of employed workers by $\ell$. 
We then define the productivity per worker $a$, the employment rate $\lambda$ and the nominal wage rate as

\begin{equation}
\label{eq:productivity}
a =\frac{Y}{\ell}, \qquad
\lambda  = \frac{\ell}{N} = \frac{Y}{aN}, \qquad
\mathrm{w}=\frac{W}{\ell},
\end{equation}
whereas the unit cost of production, defined as the wage bill divided by quantity produced, is given by
\begin{equation}
\label{eq:unit_cost}
c=\frac{W}{Y}=\frac{\mathrm{w}}{a}.
\end{equation}
We will assume throughout that productivity and workforce grow exogenously according to the dynamics 

\begin{equation}
\label{eq:productivity and population}
\frac{\dot a}{a}  = \alpha, \qquad
\frac{\dot N}{N} = \beta.  
\end{equation} 

\paragraph{Nominal quantities}
Denoting the unit price level for the homogenous good by $p$, it is clear that the portion of 
the nominal output consisting of sales should be given by $pY^d=pC+pI_k$. Accounting for inventory changes 
is less straightforward, as there can be many alternative definitions of the cost of inventory. 
We follow \cite{GodleyLavoie2007} and value inventory changes at cost $c$, 
as this is what is incurred by the acquisitions department of a firm (represented by the term $-c\dot V$ under 
the capital account in Table \ref{table}) in order to purchase unsold goods from the production department (represented 
by a revenue $+c\dot V$ in the current account). 

We therefore find that nominal output is given by
\footnote{This corresponds to equations (8.24) and (8.25) in \cite{GodleyLavoie2007}.}
\begin{equation}
\label{nominal}
Y_n = pC+pI_k+c\dot{V}= pY_d+c\dot V= pY_d + \frac{d(cV)}{dt}-\dot{c}V .
\end{equation} 
In other words, 
nominal output consists of nominal sales plus change in value of inventory 
minus an {\em inventory value adjustment} term $\dot{c}V$.
It is important to emphasize that even though total output in real terms satisfies $Y=Y_d+\dot{V}$ according to \eqref{eq:inventory}, 
nominal output is given by $Y_n = pY_d+c\dot{V}$. 
In other words, the relationship $Y_n=pY$ is true if and only if either $p=c$ or $\dot{V}=0$.

\paragraph{Financial Balances}
Further denoting household deposits by $M$ and loans to firms by $D$, 
we arrive at the balance sheet, transactions, and flow of funds described in Table \ref{table} for this economy. Notice
that we are assuming, for simplicity, that households do not borrow from banks and firms do not keep 
positive deposits, preferring to use any balances to repay their loans instead. 

\begin{table*}[t]
	\centering
			\begin{tabular}{|l|c|cc|c|c|}
				\hline
				& Households & \multicolumn{2}{|c|}{Firms} & Banks &  Sum  \\
				\hline 
				{\bf Balance Sheet} &  & &  & &  \\
				Capital stock &  & \multicolumn{2}{|c|}{$+pK$} &   & $+pK$ \\
				Inventory & & \multicolumn{2}{|c|}{$+cV$} &   & $+cV$ \\
				Deposits & $+M$ & \multicolumn{2}{|c|}{} &  $-M$&  0  \\
				Loans &  & \multicolumn{2}{|c|}{$-D$} & $+D$ &   0 \\
				\hline
				Sum (net worth) & $X_h$ & \multicolumn{2}{|c|}{$X_f$}  & $X_b$  & $X$ \\
				\hline 
				\hline
				{\bf Transactions} & &  current & capital &    &\\
				Consumption  & $-pC_h$ & $+pC$ & & $-pC_b$   &  0 \\
				Capital Investment  & & $+pI_k$ & $-pI_k$ &   & 0  \\
				Change in Inventory & & $+c\dot{V}$ & $-c\dot{V}$ & & 0 \\
				Accounting memo [GDP] & & [$Y_n$]  & &  & \\
				Wages & $+W$ & $-W$ & &   & 0 \\
				Depreciation & &$-p\delta K$ &$+p\delta K$ & & 0 \\
				Interest on deposits  & $+r_{m}M$ &  & &   $-r_{m}M$ &   0 \\
				Interest on loans  &  & $-rD$ &  & $+rD$ &  0 \\
				Profits &  & $-\Pi$ & $+\Pi$ & &     0\\
				\hline
				Financial Balances & $S_h$ & 0 & $S_f-p(I_k-\delta K)-c\dot{V}$ & $S_b$  & 0 \\
				\hline
				\hline
				{\bf Flow of Funds} & &  &    &    &\\
				Change in Capital Stock & & \multicolumn{2}{|c|}{$+p(I_k-\delta K)$}& & $+p(I_k-\delta K)$\\
				Change in Inventory & & \multicolumn{2}{|c|}{$+c\dot{V}$}& & $+c\dot{V}$\\
				Change in Deposits & $+\dot{M}$& \multicolumn{2}{|c|}{}&  $-\dot{M}$&   0  \\
				Change in Loans & & \multicolumn{2}{|c|}{$-\dot{D}$}   &  $+\dot{D}$ &   0 \\
				\hline
				Column sum & $S_h$ &  \multicolumn{2}{|c|}{$S_f$}  &   $S_b$ & $pI_k+c\dot{V}$\\
				Change in net worth & $\dot X_h=S_h$ &\multicolumn{2}{|c|}{$\dot X_f=S_f+\dot{p}K+\dot cV$}   &$\dot X_b=S_b$ & $\dot X$\\
				\hline
			\end{tabular}
		\caption{Balance sheet and transactions flows.}
		\label{table}
\end{table*}%

It follows from Table \ref{table} that the net profit for firms, after paying wages, interest on debt, and accounting 
for depreciation (i.e consumption of fixed capital) is given by
\begin{align*}
\Pi&= Y_n - W -rD - p\delta K.
\end{align*}
Observe that, even though changes in inventory add to profits, 
sales constitute the only way for firms to have positive gross profits (i.e before interest and depreciation)
since  
\begin{equation*}
Y_n-W=pC+pI_k+c\dot{V} -W = pY_d+c(\dot{V}-Y)=(p-c)Y_d.
\end{equation*}

It is also assumed in Table \ref{table} that all profits are reinvested, that is, $\Pi=S_f$. The financial balances 
row on Table \ref{table} corresponds to the following {\it ex post} accounting identity 
between total nominal savings and investment in the economy:
\begin{equation}
S = S_h+S_f+S_b = p(I_k-\delta K) +c \dot{V} \: .
\end{equation}

In particular for the firm sector we have 
\begin{equation}
\label{debt_franke}
\dot D = p(I_k-\delta K)+c\dot V -\Pi = pI_k+c\dot V - \Pi_p,
\end{equation}
where $\Pi_p=Y_n-W-rD$ denotes the pre-depreciation profit. 

\paragraph{Intensive variables} To obtain a steady state in a growing economy, we normalize real variables by dividing 
them by total output $Y$, namely
\begin{equation}
\label{eq:real state variables}
y_e=\frac{Y_e}{Y}, \quad y_d=\frac{Y_d}{Y}, \quad \v=\frac{V}{Y},
\end{equation}
and nominal variables by dividing them by $pY$ (even though this 
is not equal to nominal output according to the remark following \eqref{nominal}), that is
\begin{equation}
\label{eq:franke wage share}
\omega = \frac{W}{pY}=\frac{c}{p}=\frac{{\rm w}}{pa} \; , \quad d= \frac{D}{pY}.
\end{equation}
We use $\omega$ and $d$ as proxies for the {\em actual} wage share $W/Y_n$ and debt ratio $D/Y_n$, which case can readily be obtained by dividing 
the expressions above by 
\begin{equation}
\frac{Y_n}{pY}=\frac{pY_d+c(Y-Y_d)}{pY}=\frac{(p-c)Y_d}{pY}+\frac{c}{p}=(1-\omega)y_d+\omega.
\end{equation}

\subsection{Behavioural Rules}
\label{sec:dynamical model}

We now specify the behavioural rules for firms, banks, and households. 
Namely, for given values of the state variables, 
firms decide the level of capital investment $I_k$, 
planned changes in inventory $I_p$, 
and expected sales $Y_e$, 
whereas banks and households decide the level of consumption $C_b$ and $C_h$. 
This in turn determines capital by \eqref{eq:capital}, 
output by \eqref{eq:supply}, utilization by \eqref{eq:utilization}, 
sales demand by \eqref{eq:demand}, 
and unplanned changes in inventory by \eqref{eq:unplanned}. 
Consequently, 
since productivity and workforce growth are exogenous, 
the level of output $Y$ in turn gives the number of employed workers $\ell$ 
and the employment rate $\lambda$ by \eqref{eq:productivity}. 
Further specification of the dynamics for the nominal wage rate ${\mathrm w}$ and prices $p$ then completes a model.  

\paragraph{Firms}
We start by assuming that firms forecast the long-run growth rate of the economy to be 
a function $g_e(u,\pi_e)$ of utilization $u$ and (pre-depreciation) {\em expected} profitability $\pi_e$ defined as 
\begin{equation}
\label{eq:franke expected profit}
\pi_e = \frac{Y_{ne}-W-rD}{pY},
\end{equation}
where $Y_{ne}=pY_e+cI_p$ denotes the expected nominal output. Inserting \eqref{eq:real state variables} and \eqref{eq:franke wage share} and into \eqref{eq:franke expected profit}, we find that the expected profitability can be expressed as
\begin{equation}
\label{eq:franke expected profit share}
\pi_e = y_e(1-\omega) - rd\;.
\end{equation}

In addition to taking into account the long-run growth rate $g_e(u,\pi_e)$, firms adjust their short-term expectations based to the observed level of demand.
This leads to the following dynamics for expected demand:
\begin{equation}
\label{expected_sales_franke}
\dot Y_e= g_e(u,\pi_e) Y_e+\eta_e(Y_d-Y_e) \;, 
\end{equation}
for a constant $\eta_e \geq 0$, representing the speed of short-term adjustments to observed demand.
We assume further that firms aim to maintain inventories at a desired level 

\begin{equation}
\label{desired_inventory}
V_d=f_d Y_e \;,
\end{equation} 
for a fixed proportion $0 \leq f_d \leq 1$. 
While this means that the long-term growth rate of desired inventory level should also be $g_e(u,\pi_e)$, 
we assume again that firms adjust their short-term expectations based on the observed level of inventory.
This leads to the following expression for planned changes in inventory:
\begin{equation}
\label{planned_inventory_change}
I_p=g_e(u,\pi_e) V_d+\eta_d(V_d-V), 
\end{equation}
for a constant $\eta_d \geq 0$, representing the speed of short-term adjustments to observed inventory.

To complete the specification of firm behavior, we assume that investment is given by 
\begin{equation}
\label{eq:investment}
	I_k=\frac{\kappa(u,\pi_e)}{\nu} K\;,
\end{equation}
for a function $\kappa(\cdot,\cdot)$ capturing explicitly the effects of both capacity utilization and expected profits. Based on \eqref{eq:capital}, this leads to the following dynamics for capital,
\begin{equation}
	\label{capital_franke}
	\frac{\dot K}{K}  =\frac{\kappa(u,\pi_e)}{\nu}-\delta(u).
\end{equation}



\paragraph{Banks and Households} We assume that total consumption is given by 
\begin{equation}
\label{eq:consumption}
C = \theta(\omega,d)Y,
\end{equation}
for a function $\theta$ of the wage and debt ratios $\omega$ and $d$. This includes, for example, 
the usual case where nominal consumption of households and banks is assumed to be given by constant fractions of income and wealth, namely, 
\begin{align}
\label{consumption_1}
pC_h &=c_{ih}[W+r_{m}M]+c_{wh}M, \\
\label{consumption_2}
pC_b &=c_{ib}[rD-r_{m}M]+c_{wb}(D-M). 
\end{align}
Under the additional simplifying assumption that $c_{ih}=c_{ib}=c_i$ and $c_{wh}=c_{wb}=c_w$, we have
\begin{equation}
\label{ex:consumption}
pC=c_1W+c_2D,
\end{equation}
with $c_1= c_i$ and $c_2=c_w+c_ir$. Alternatively, we can follow \cite{Ryoo2013} and assume that $D=M_h$ and $r=r_{M_h}$, that is, 
banks have zero net worth and charge zero intermediation costs (and therefore have no consumption), 
in which case \eqref{ex:consumption} holds with $c_1=c_{ih}$ and $c_2=c_{wh}+rc_{ih}$. In either case, we see that \eqref{ex:consumption} is
an example of \eqref{eq:consumption} with 
\begin{equation}
\label{consumption_common}
\theta(\omega,d)=c_1\omega+c_2 d, 
\end{equation}
for non-negative constants $c_1$ and $c_2$. We then find that nominal demand is given by
\begin{align}
\label{eq:Y_d}
pY_d&=pC+pI_k=p\theta(\omega,d)Y+p\frac{\kappa(u,\pi_e)}{\nu}K,
\end{align}
from which we obtain the auxiliary variable 
\begin{equation}
\label{eq:demand share} 
y_d=y_d(\omega,d,y_e,u)=\frac{Y_d}{Y}=\theta(\omega,d) +\frac{\kappa(u,\pi_e)}{u}.
\end{equation}

\paragraph{Price and wage dynamics}
For the price dynamics we assume that the long-run equilibrium price is given by a constant markup $m\ge 1$ times unit labor cost $c$, 
whereas observe prices converge to this through a lagged adjustment with speed $\eta_p>0$. A second component with adjustment speed $\eta_q>0$ is added to that dynamics to take into account short-term considerations regarding unplanned changes in inventory volumes: 
\begin{equation}
\frac{\dot{p}}{p} = \eta_p \left(m \frac{c}{p} - 1\right) - \eta_{q}\frac{Y_e-Y_d}{Y}  =\eta_p \left(m\omega - 1\right) + \eta_{q}(y_d-y_e):= i(\omega,y_d,y_e) \label{inflation}.
\end{equation}
We assume that the wage rate $\w$ follows 
\begin{equation}
\label{eq:wage 3}
\frac{\dot{\w}}{\w} = \Phi(\lambda) + \gamma \frac{\dot p}{p},
\end{equation}
for a constant $0\leq\gamma\leq 1$. The assumption states that workers bargain for wages based on the current state of the labour market, 
but also take into account the observed inflation rates. 
The constant $\gamma$ represents the degree of money illusion, 
with $\gamma=1$ corresponding to the case where workers fully incorporate inflation in their bargaining.

\section{The main dynamical system}
\label{sec:main}

Combining \eqref{desired_inventory} and \eqref{planned_inventory_change}, we
see that output is given by 
\begin{equation}
\label{output_franke}
Y=Y_e+I_p =\big[f_d(g_e(u,\pi_e)+\eta_d)+1\big]Y_e-\eta_d V \;,
\end{equation}
so that the inventory-to-output ratio $\v$ is given by
\begin{equation}
\label{eq:v}
\v = \frac{V}{Y} = \frac{[1+f_d (g_e(u,\pi_e)+\eta_d)]y_e - 1}{\eta_d} \;.
\end{equation}
Differentiating \eqref{output_franke} and using \eqref{expected_sales_franke} and \eqref{eq:inventory}, we obtain the following dynamics for 
output 
\begin{equation}
\frac{\dot Y}{Y}= \big[f_d(g_e(u,\pi_e)+\eta_d)+1\big]\big(y_eg_e(u,\pi_e)+\eta_e(y_d-y_e)\big)+\eta_d(y_d-1) =: g(u,\pi_e,y_d,y_e) \label{output_dynamics_franke}  
\end{equation}

\label{sec:reduced dynamical system}

The dynamics for the wage share $\omega=\rm{w}/(pa)$ follows from \eqref{eq:wage 3} and \eqref{inflation}:
\begin{equation}
\label{wage_keen_modified}
\frac{\dot\omega}{\omega}= \frac{\dot{\mathrm{w}}}{\mathrm{w}}-\frac{\dot a}{a}-\frac{\dot p}{p}=\Phi(\lambda)-\alpha-(1-\gamma)i(\omega,y_d,y_e),
\end{equation}
where the inflation rate is defined in \eqref{inflation}. For the employment rate $\lambda=Y/(aN)$, we use \eqref{output_dynamics_franke} and \eqref{eq:productivity and population} to obtain 
\begin{equation}
\label{employment_dynamics_franke}
\frac{\dot{\lambda}}{\lambda} = \frac{\dot{Y}}{Y} - \frac{\dot a}{a}-\frac{\dot N}{N}= g(u,\pi_e,y_d,y_e)-\alpha-\beta \;.
\end{equation}
For the debt ratio $d=D/(pY)$, using the expression for debt change in \eqref{debt_franke}, we find that
\begin{align}
\frac{\dot d}{d} =\left[\frac{\dot D}{D}-\frac{\dot p}{p}-\frac{\dot Y}{Y}\right]&=\left[\frac{pI_k+c\dot V-(Y_n-W-rD)}{D}\right] -i(\omega, y_d,y_e)-g(u,\pi_e,y_d,y_e)\nonumber \\
&=\left[\frac{W+rD-pC)}{D}\right] -i(\omega,y_d,y_e)-g(u,\pi_e,y_d,y_e)\nonumber \\
 \label{eq:d}&=\left[\frac{\omega+rd-\theta(\omega,d))}{d}\right] -i(\omega, y_d,y_e)-g(u,\pi_e,y_d,y_e).
\end{align}
Similarly, for the expected sales ratio $y_e=Y_e/Y$, we use \eqref{expected_sales_franke} to obtain
\begin{align}
\label{eq:dot y_e}
\frac{\dot y_e}{y_e}&=\frac{\dot Y_e}{Y_e}-\frac{\dot Y}{Y}=g_e(u,\pi_e)+\eta_e\left(\frac{y_d}{y_e}-1\right)-g(u,\pi_e,y_d,y_e).
\end{align}
Finally, for the capacity utilization $u=\nu Y/K$, using \eqref{capital_franke} we find
\begin{equation}
\frac{\dot u}{u}=\frac{\dot Y}{Y}-\frac{\dot K}{K}=g(u,\pi_e,y_d,y_e)-\frac{\kappa(u,\pi_e)}{\nu}+\delta(u).
\end{equation}

Since $y_d$ is expressed in \eqref{eq:demand share} as a function of $(\omega, d, \pi_e, u)$ and  
and $\pi_e$ is given in \eqref{eq:franke expected profit share} as a function of $(\omega, d, y_e)$, we see 
that the model can be completely characterized by the state variables $(\omega, \lambda, d, y_e, u)$ 
satisfying the following system of ordinary differential equations:
\begin{equation}
\label{franke}
\left\{
\begin{array}{ll}
\dot\omega &= \omega\left[\Phi(\lambda)-\alpha-(1-\gamma)i(\omega, y_d,y_e)\right] \\
\dot\lambda  &= \lambda\left[g(u,\pi_e,y_d,y_e)-\alpha-\beta\right]  \\
\dot d   &= d \left[ r-g(u,\pi_e,y_d,y_e)- i(\omega, y_d,y_e)\right] +\omega- \theta(\omega,d) \\
\dot y_e  &=y_e\left[g_e(u,\pi_e)-g(u,\pi_e,y_d,y_e)\right]+\eta_e(y_d-y_e) \\
\dot u  &=u\left[g(u,\pi_e,y_d,y_e)-\frac{\kappa(u,\pi_e)}{\nu}+\delta(u)\right]
\end{array}
\right.
\end{equation}
where $i(\omega, y_d,y_e)$ is given by \eqref{inflation} and $g(u,\pi_e,y_d,y_e)$ is given by \eqref{output_dynamics_franke}.


To obtain an interior equilibrium point $(\overline\omega,\overline\lambda,\overline d,\overline y_e, \overline u)$, observe that the second equation in \eqref{franke} requires that
\begin{equation}
\label{eq:franke equilibrium condition 1}
g(\overline u,\overline\pi_e,\overline y_d,\overline y_e) = \alpha+\beta\;,
\end{equation}
which when inserted in the forth equation leads to $\overline y_d=\overline y_e$ and
\begin{equation}
g_e(\overline u,\overline\pi_e)=\alpha+\beta
\end{equation}
at equilibrium. Using this and \eqref{eq:franke equilibrium condition 1} in 
\eqref{output_dynamics_franke} therefore gives 
\begin{equation}
\label{y_bar_franke}
\overline y_d=\overline y_e=\frac{1}{1+(\alpha+\beta)f_d} \;.
\end{equation}
Inserting \eqref{y_bar_franke} into \eqref{eq:v} implies that $\overline \v=f_d \overline y_e$, so that 
the equilibrium level of inventory is the desired level $V_d=f_d\overline y_e Y$. 
Substituting $\overline y_d=\overline y_e$ into \eqref{inflation} leads to an equilibrium inflation of the form 
\begin{equation}
\label{eq_inflation}
i(\overline\omega, \overline y_d, \overline y_e) = i(\overline\omega)=\eta_p(m\overline\omega-1),
\end{equation} 
that is, without any inventory effects. Using the third equation in \eqref{franke} we see that the debt ratio at equilibrium satisfies 
\begin{equation}
\label{d_eq}
\overline d=\frac{\overline \omega-\theta(\overline\omega,\overline d)}{\alpha+\beta+i(\overline\omega)-r}.
\end{equation}
Moving to the last equation in \eqref{franke}, we obtain that the investment function at equilibrium satisfies 
\begin{equation}
\label{kappa_eq}
\kappa(\overline\pi_e,\overline u)= \nu[\alpha+\beta+\delta(\overline u)],
\end{equation}
which can be inserted in \eqref{eq:demand share} to yield the equilibrium capacity utilization as the solution to 
\begin{equation}
\label{eq:equilibrium u as a function}
\overline u = \frac{\nu[\alpha+\beta+\delta(\overline u)](1+(\alpha+\beta)f_d)}{1- \theta(\overline\omega,\overline d)(1+(\alpha+\beta)f_d)}\;.
\end{equation}
We can then obtain the values of $(\overline\omega,\overline d)$ by solving \eqref{d_eq}-\eqref{kappa_eq} 
with $\overline\pi_e$ defined from \eqref{eq:franke expected profit share}. Finally, returning to the first equation in \eqref{franke} we find the equilibrium employment rate by solving
\begin{equation}
\label{eq:lambda LR}
\Phi(\overline\lambda) = \alpha+ (1-\gamma)i(\overline\omega).
\end{equation}
We therefore see that existence and uniqueness of the interior equilibrium depends on 
properties of the functions $\kappa$ and $\theta$, which need to be asserted in specific realizations of the model. 

To summarize, an interior equilibrium of \eqref{franke} is characterized by a constant 
growth rate of output equal to $\alpha+\beta$, constant capacity utilization, expected sales equal to demand, and the level 
of inventory equal to a constant proportion of expected sale. The present model is nevertheless highly complex.
It needs the specification of at least fourteen parameters in addition to three behavioural functions. 
The exploration of other possible equilibrium points is considerably involved,
and any local stability analysis will reveal to be a cumbersome and non-intuitive exercise.

In order to build intuition about the system, we follow the strategy of considering the lower-dimensional subsystems 
that arise in some limiting cases for the model parameters and behavioural functions. We start with a few special cases corresponding to known models in the 
literature. 

\subsection{The Goodwin model}
\label{goodwin_section}

The simplest special case of \eqref{franke} consists of the model proposed in \cite{Goodwin1967}. The original Goodwin model is 
formulated in real terms, which we can easily reproduce by setting $\eta_p=\eta_q=\gamma=0$, meaning that the rate of inflation is zero, 
and setting $p=1$. It also makes no reference to inventories, thereby implicitly assuming that output equals demand. 
We can recover this from the general model of the previous section by assuming that $f_d=\eta_d=0$, 
meaning that there is no desired inventory level ($V_d=0$) or planned investment in inventory ($I_p=0$), and that $\eta_e\to\infty$, meaning that firms 
have perfect forecast of demand and set $Y_e=Y_d=Y$ at all times. In addition, Goodwin 
adopts a constant capital-to-output ratio, which we can recover by setting $u=1$. Finally, although not explicitly mentioned in \cite{Goodwin1967}, 
we adopt a constant depreciation rate $\delta(u)=\delta>0$ for the Goodwin model. 

The only explicit assumption of the Goodwin model regarding the behaviour of firms is that
investment is equal to profits, which in the present setting corresponds to 
\[\kappa(u,\pi_e)=\pi_e=1-\omega-rd,\]
since $y_e=Y_e/Y=1$ in \eqref{eq:franke expected profit share}. The model is also silent about banks, but it follows
from \eqref{debt_franke} and the investment rule above (recalling that $\dot V=0$) that $\dot D=0$ at all times, so we assume for simplicity that 
$d=D_0=0$. Alternatively we could adopt an arbitrary constant level of debt $D_0$, observing that, in a growing economy,  
$d=D_0/Y\to 0$.

Regarding households, the assumption in \cite{Goodwin1967} is that all wages are consumed, namely $c_{ih}=c_1=1$ 
in the notation of \eqref{consumption_1}. For consistency, we set $c_2=r$, even though this is not relevant when $D=0$. 

For the growth rate, observe that we can no longer obtain it by simply differentiating \eqref{output_franke}, since \eqref{expected_sales_franke} is degenerate in the limit case $\eta_e\to\infty$. Instead, since $u=1$, we can use the fact that $Y=\frac{K}{\nu}$ to obtain 
\begin{equation}
\frac{\dot Y}{Y}=\frac{\dot K}{K}=\frac{1-\omega}{\nu}-\delta.
\end{equation}

With these parameter choices, the system \eqref{franke} reduces to the familiar form 
\begin{equation}
\label{goodwin}
\left\{
\begin{array}{ll}
\dot\omega &= \omega\left[\Phi(\lambda)-\alpha\right] \\
\dot\lambda  &= \lambda\left[\frac{1-\omega}{\nu}-\alpha-\beta-\delta\right],  \\
\end{array}
\right.
\end{equation}
discussed, for example, in \cite{GrasselliCostaLima2012}. The solutions of \eqref{goodwin} are closed periodic orbits around the non-hyperbolic 
equilibrium point 
\begin{equation}
\overline \omega = 1-\nu(\alpha+\beta+\gamma), \quad \overline \lambda = \Phi^{-1}(\alpha),
\end{equation}
which we recognize as special cases of \eqref{kappa_eq} and \eqref{eq:lambda LR}, respectively. 

\subsection{The Franke model}
\label{franke_section}

As mentioned in Section \ref{sec:introduction}, our proposed dynamics for inventories follows closely the Metzlerian model formalized 
in \cite{Franke1996}. The Franke model is also formulated in real terms, so we maintain the choice of 
$\eta_p=\eta_q=\gamma=0$ and $p=1$ from the previous section, and normalizes all variables by dividing them 
by $K$ instead of $Y$, resulting in the intensive variables 
\begin{equation*}
u^F:=Y/K=u/\nu, \quad z^F:=Y_e/K=y_eu^F, \quad \v^F=:V/K=\v u^F.
\end{equation*}
Crucially, the model in \cite{Franke1996} implicitly assumes a constant wage share $\omega$ (see, for example, footnote 9 on 
page 246), so that the first equation in \eqref{franke} is simply $\dot \omega = 0$. The second equation in \eqref{franke} 
then decouples from the rest of the system and simply provides the employment rate along the solution path, in particular 
leading to a constant employment rate at equilibrium. As with the Goodwin model, the Franke model is also silent about banks, implicitly 
assuming that firms can raise the necessary funds for investment through retained profits and savings from households, which we 
reproduce here by setting $\dot d=0$ in \eqref{franke}. 

The behaviour of firms, on the other hand, is almost identical to the one adopted here, 
with our equations \eqref{expected_sales_franke}, \eqref{desired_inventory}, and 
\eqref{planned_inventory_change} corresponding directly to equations (7), (2), and (3) in \cite{Franke1996}, respectively, provided 
we take
\begin{equation}
\label{growth_franke}
g_e(u,\pi_e)=\alpha+\beta
\end{equation}
as the long-run growth rate of expected sales. For 
the investment function, we recover the assumption in \cite{Franke1996} by setting 
\begin{equation}
\kappa(u,\pi_e)=\nu h(u^F), 
\end{equation}
for an increasing function $h(\cdot)$. Regarding effective demand, instead of modelling consumption and investment separately, the assumption 
in \cite{Franke1996} is that demand in excess of output is given directly in terms of utilization, which we can reproduce in our model by setting 
\begin{equation}
y^d=e(u^F)+1,
\end{equation}
for a decreasing function $e(\cdot)$. With these choices, it is a simple exercise to verify that the fourth and fifth equations in \eqref{franke} are equivalent to equations (9)-(10) 
for $\v^F$ and $z^F$ in \cite{Franke1996}, with equilibrium values given by 
\begin{equation}
\overline \v^F = \frac{f_d \overline u^F}{1+(\alpha+\beta)f_d}=\overline v \, \overline u^F, \qquad \overline z^F = \frac{\overline u^F}{1+(\alpha+\beta)f_d}=\overline y_e \overline u^F.
\end{equation}
It is shown in \cite{Franke1996} that this equilibrium is locally asymptotically stable provided the speed of adjustment of inventories $\eta_d$ is sufficiently 
small. For $\eta_d$ above a certain threshold, however, local stability can only be asserted when the speed of adjustment of expected sales $\eta_e$ is
sufficiently small. The main innovation in \cite{Franke1996} consists of adopting a variable speed of adjustment  $\eta_d=\eta_d(z^F)$ and 
investigate its effects on the stability of the equilibrium. It is then shown that even in the unstable case, namely when both $\eta_d(\overline z^F)$ and 
$\eta_e$ are large enough so that the equilibrium is locally repelling, global stability can be achieved provided $\eta_d(z^F)$ decreases fast enough away 
from the equilibrium. As stated in \cite{Franke1996}, the equilibrium is ``locally repelling, but it is attractive in the outer regions of the state space'', giving rise 
to periodic orbits. 

\subsection{The original Keen model}
\label{keen_section}
The model proposed in \cite{Keen1995} is based on the same assumptions of the Goodwin model regarding the price-wage dynamics ($\eta_p=\eta_q=\gamma=0$ and $p=1$), desired inventory level ($f_d=\eta_d=V_d=I_p=0$), expected sales ($\eta_e\to\infty$ and $Y_e=Y_d=Y$), constant capital-to-output ratio with 
full utilization ($u=1$), and constant depreciation rate ($\delta(u)=\delta$). The innovation in the 
model is that investment is now given by 
\begin{equation}
\label{investment_keen}
\kappa(u,\pi_e)=\kappa(\pi_e)=\kappa(1-\omega-rd),
\end{equation}
where we used the fact that $y_e=Y_e/Y=1$ in \eqref{eq:franke expected profit share}. Moreover, the identity $Y_d=Y$ implies that 
\begin{equation}
\label{consumption_keen}
C=Y_d-I_k=(1-\kappa(\pi_e))Y,
\end{equation}
that is, $\theta(\omega,d)=1-\kappa(1-\omega-rd)$ in \eqref{eq:consumption}. In other words, in the absence of either price or quantity adjustments, total consumption plays the role of an accommodating variable in the model. 

Since \eqref{expected_sales_franke} is degenerate in the limit case $\eta_e\to\infty$, we again use $Y=\frac{K}{\nu}$ 
instead of \eqref{output_franke} to obtain the growth rate of the economy as
\begin{equation}
\label{growth_keen}
\frac{\dot Y}{Y}=\frac{\dot K}{K}=\frac{\kappa(\pi_e)}{\nu}-\delta.
\end{equation}

With these parameter choices, the system \eqref{franke} reduces to
\begin{equation}
\label{keen}
\left\{
\begin{array}{ll}
\dot\omega &= \omega\left[\Phi(\lambda)-\alpha\right] \\
\dot\lambda  &= \lambda\left[\frac{\kappa(\pi_e)}{\nu}-\alpha-\beta-\delta\right]  \\
\dot d   &= d \left[r-\frac{\kappa(\pi_e)}{\nu}-\delta\right] +\omega -1+\kappa(\pi_e)\\
\end{array}
\right.
\end{equation}
where $\pi_e=1-\omega-rd$. It is then easy to see that \eqref{d_eq}, \eqref{kappa_eq} and \eqref{eq:lambda LR} reduce to 
\begin{equation}
\overline d =\frac{\overline \omega-1+\nu(\alpha+\beta+\delta)}{\alpha+\beta-r},\quad \kappa(\overline \pi_e) = \nu(\alpha+\beta+\delta), \quad  
\Phi(\overline \lambda)=\alpha,
\end{equation}
from where we obtain the interior equilibrium point $(\overline\omega_1,\overline\lambda_1,\overline d_1)$ found in \cite{GrasselliCostaLima2012}, 
which is shown to be locally stable provided the investment function $\kappa(\cdot)$ is sufficiently increasing at equilibrium, but does not exceed the 
amount of net profits by too much. 

Apart from the interior equilibrium, \cite{GrasselliCostaLima2012} established that the system \eqref{keen} admits an equilibrium characterized by 
$(\overline\omega_2,\overline\lambda_2,\overline d_2)=(0,0,+\infty)$, which is locally asymptotically stable provided 
\begin{equation}
\lim_{\pi_e\to-\infty}\kappa(\pi_e)<\nu(r+\delta)
\end{equation}
a condition that is likely to be satisfied for typical parameters. 

\subsection{Monetary Keen model}
\label{keen_inflation_section}

As shown in \cite{GrasselliNguyenHuu2015}, it is relatively straightforward to incorporate the price-wage dynamics in \eqref{inflation}-\eqref{eq:wage 3} in
the original Keen model. Adopting all the parameter choices and functional forms of the previous section (including $\eta_q=0$) with the exception of arbitrary 
constants $\eta_p$ and $\gamma$, we find that \eqref{franke} reduces to the three-dimensional system 
\begin{equation}
\label{keen_inflation}
\left\{
\begin{array}{ll}
\dot\omega &= \omega\left[\Phi(\lambda)-\alpha-(1-\gamma)i(\omega)\right] \\
\dot\lambda  &= \lambda\left[\frac{\kappa(\pi_e)}{\nu}-\alpha-\beta-\delta\right]  \\
\dot d   &= d \left[r-\frac{\kappa(\pi_e)}{\nu}-\delta-i(\omega)\right] +\omega -1+\kappa(\pi_e)\\
\end{array}
\right.
\end{equation}
where $\pi_e=1-\omega-rd$ and $i(\omega)=\eta_p(m\omega-1)$. Solving \eqref{d_eq}, \eqref{kappa_eq} and \eqref{eq:lambda LR} for this system 
gives an interior equilibrium $(\overline\omega_1,\overline\lambda_1,\overline d_1)$ analogous to that of the original Keen model. Apart 
from it, \cite{GrasselliNguyenHuu2015} showed that the system \eqref{keen_inflation} also admits the equilibrium 
$(\overline\omega_2,\overline\lambda_2,\overline d_2)=(0,0,+\infty)$, as well as a new class equilibria of the form 
$(\overline\omega_3,0,\overline d_3)$ or $(\overline\omega_3,0,+\infty)$ where 
\begin{equation}
\overline\omega_3=\frac{1}{m}+\frac{\Phi(0)-\alpha}{m\eta_p(1-\gamma)}
\end{equation}
is a wage share satisfying $i(\overline\omega_3)<0$ (deflation) and $\overline d_3$ is a finite debt ratio obtained as 
the solution of a non-linear equation. The stability of all three types of equilibrium is analyzed in detail in \cite{GrasselliNguyenHuu2015}, with 
the overall conclusion that ``money emphasizes the stable nature of asymptotic states of the economy, both desirable and undesirable."

%

\section{Long-run dynamics}
\label{sec:effective demand in the keen model}

As mentioned in Section \ref{sec:dynamical model}, the core dynamics for expected sales and inventories in the model consists of the interplay 
between long-run expectations and short-run fluctuations characterized by equations \eqref{expected_sales_franke}-\eqref{planned_inventory_change}. It 
is therefore instructive to investigate the properties of the sub-models that arise when each of these effects is considered separately. We start with 
case where short-run fluctuations are ignored by firms, namely when $\eta_e=\eta_d=0$. In addition, we assume that there are no planned 
changes in inventories, namely that $f_d=0$, so that $Y=Y_e$ and any discrepancy between supply and demand is absorbed by unplanned 
inventory changes $\dot V=Y-Y_d$. 

Observe that, in this case, the growth rate of the economy is given by 
\begin{equation}
g(u,\pi_e,y_d,y_e) = g_e(u,\pi_e),
\end{equation}
as there is no feedback channel from the demand $y_d$ on either 
expected sales $Y_e$ or planned inventories $I_p$, and consequently no impact of demand on output $Y$, which is therefore 
solely determined by the expected long-run growth rate of the economy. Consequently, the fourth equation in \eqref{franke} is identically zero, 
which is consistent with the fact that $y_e=Y_e/Y=1$, and the system reduces to 
\begin{equation}
\label{keen_demand}
\left\{
\begin{array}{ll}
\dot\omega &= \omega\left[\Phi(\lambda)-\alpha-(1-\gamma)i(\omega,y_d)\right] \\
\dot\lambda  &= \lambda\left[g_e(u,\pi_e)-\alpha-\beta\right]  \\
\dot d   &= d \left[r-g_e(u,\pi_e)-i(\omega,y_d)\right] +\omega-\theta(\omega,d) \\
\dot u  &=u\left[g_e(u,\pi_e)-\frac{\kappa(u,\pi_e)}{\nu}+\delta(u)\right],
\end{array}
\right.
\end{equation}
where $\pi_e=1-\omega-rd$ and $i(\omega,y_d)=\eta_p(m\omega-1)+\eta_q(y_d-1)$. Notice also that the inventory-to-output ratio can no longer 
be determined by \eqref{eq:v} (since $\eta_d=0$), but should instead be found from the auxiliary equation 
\begin{equation}
\label{v_demand}
\dot{\rm{v}}=(1-y_d)-g_e(u,\pi)\rm{v}.
\end{equation}
The interior equilibrium for \eqref{keen_demand} is obtained from \eqref{d_eq}-\eqref{eq:lambda LR} with $\eta_e=\eta_d=f_d=0$. In particular, since $\overline y_d=1$, this equilibrium implies that $\rm{v}\to \overline v=0$ according to \eqref{v_demand}, which is consistent with \eqref{y_bar_franke} and $\overline v = f_d \overline y_e$ with $f_d=0$. This is the analogue of the {\em good} equilibrium for the original Keen model, corresponding to a finite debt ratio and non-zero wage share and employment rate.

Observe that in the special case $g_e(u,\pi_e)=\alpha+\beta$, that is, a constant long-run growth rate for 
expected sales corresponding to the model proposed in \cite{Franke1996}, we find that the employment 
rate in \eqref{keen_demand} is constant, as should be expected in a model where the output growth rate is 
identical to the sum of population and productivity growth rates. But this immediately implies that the interior equilibrium in \eqref{d_eq}-\eqref{eq:lambda LR} can 
only be achieved for the initial condition $\lambda_0=\Phi^{-1}(\alpha+(1-\gamma)\overline \omega)$, with any smaller initial  
employment rate leading to $\omega\to 0$ and any bigger one leading to $\omega\to\infty$. We therefore do not pursue 
this special case further. 

Alternatively, the special case 
\begin{equation}
\label{growth_keen_2}
g_e(u,\pi_e)=\frac{\kappa(u,\pi_e)}{\nu}-\delta(u)
\end{equation}
is conceptually much closer to the original Keen 
model in \cite{Keen1995}, in that expected sales (and therefore output) grow at the same rate as capital. In this case, capacity utilization is given by a constant $u_0$, as the fourth equation in \eqref{keen_demand} vanishes and the model reduces to a three-dimensional system for $(\omega,\lambda,d)$, which we now consider in its 
real and monetary versions. 

\subsection{Real version}
\label{non_monetary_long}

Consider first the model in real terms, that is to say, with the wage-price parameters set to $\eta_p=\eta_q=\gamma=0$ and $p=1$ as in the original Goodwin and Keen models. We then obtain that the system \eqref{keen_demand}, with $g_e(u,\pi_e)$ given by \eqref{growth_keen_2}, reduces to 
\begin{equation}
\label{keen_demand_2}
\left\{
\begin{array}{ll}
\dot\omega &= \omega\left[\Phi(\lambda)-\alpha\right] \\
\dot\lambda  &= \lambda\left[\frac{\kappa(u_0,\pi_e)}{\nu}-\delta(u_0)-\alpha-\beta\right]  \\
\dot d   &= d \left[r-\frac{\kappa(u_0,\pi_e)}{\nu}+\delta(u_0)\right] +(1-c_1)\omega-c_2d, \\
\end{array}
\right.
\end{equation}
where $\pi_e=1-\omega-rd$ as before, and we have adopted a consumption function of the form $\theta(\omega,d)=c_1\omega+c_2d$. We regard this as the closest model to the original Keen model in \eqref{keen}, but 
with a non-trivial effective demand of the form 
\begin{equation}
\label{effective_demand_keen}
y_d=c_1\omega+c_2d+\frac{\kappa(u_0,\pi_e)}{u_0}
\end{equation}
and fluctuating inventory levels given by 
\begin{equation}
\label{v_keen_demand}
\dot{\rm{v}}=\left(1-c_1\omega-c_2d-\frac{\kappa(u_0,\pi_e)}{u_0}\right)-\left(\frac{\kappa(u_0,\pi_e)}{\nu}-\delta(u_0)\right)\rm{v}.
\end{equation}

The system \eqref{keen_demand_2} also admits a {\em bad} equilibrium of the form $(\omega,\lambda,d)=(0,0,+\infty)$. Nevertheless, for  
$c_2>0$ (or any other consumption function that includes a positive wealth effect), this equilibrium implies that 
$y_d\to +\infty$ and consequently $v\to -\infty$, which is not economically meaningful. For this reason, in the
next section we investigate a monetary version of the model where inflation becomes infinite as $y_d\to +\infty$ in accordance with the price 
dynamics \eqref{inflation} with $\eta_q>0$. Interestingly, the system \eqref{keen_demand_2} also admits a bad equilibrium of the form 
$(\omega,\lambda,d)=(0,0,0)$, which was not possible in the 
original Keen model. Nevertheless, it is easy to see that this equilibrium is unstable provided 
\begin{equation}
\label{growth_1}
g_e(u_0,1)>\alpha+\beta,
\end{equation}
a condition that is likely to be satisfied in practice.

\subsection{Monetary version}

Using \eqref{inflation}-\eqref{eq:wage 3} as the price-wage dynamics leads to the 
following monetary version of the model of the previous section 
\begin{equation}
\label{keen_demand_inflation}
\left\{
\begin{array}{ll}
\dot\omega &= \omega\left[\Phi(\lambda)-\alpha-(1-\gamma)i(\omega,d)\right] \\
\dot\lambda  &= \lambda\left[\frac{\kappa(u_0,\pi_e)}{\nu}-\delta(u_0)-\alpha-\beta\right]  \\
\dot d   &= d \left[r-\frac{\kappa(u_0,\pi_e)}{\nu}+\delta(u_0)-i(\omega,d)\right] +(1-c_1)\omega-c_2d \\
\end{array}
\right.
\end{equation}
where $\pi_e=1-\omega-rd$ and 
\begin{equation}
i(\omega,d) = \eta_p \left(m\omega - 1\right) + \eta_{q}(y_d-1) = \eta_p \left(m\omega - 1\right) + \eta_{q}\left(c_1\omega+c_2d+\frac{\kappa(u_0,\pi_e)}{u_0}-1\right).
\end{equation}
As before, we regard this as the closest model to the monetary Keen model in \eqref{keen_inflation}, but with a non-trivial effective demand given by \eqref{effective_demand_keen} and fluctuating inventory levels given by \eqref{v_keen_demand}. 

In what follows, we denote $\delta(u_0)=\delta>0$ 
and $\kappa(u_0,\pi_e)=\kappa(\pi_e)$ for convenience. Let $ J(\omega, \lambda, d)$ be the Jacobian matrix of \eqref{keen_demand_inflation}, given by
\begin{equation}
\label{eq:jacobian_keen_modified}
\left[
\begin{array}{ccc}
J_{1,1}(\omega,d)	& \omega \Phi'(\lambda)				& J_{1,3}(\omega,d) \\
-\dfrac{\lambda}{\nu}\kappa'(\pi_e)			& \dfrac{\kappa(\pi_e)}{\nu}-\alpha-\beta-\delta 	& -\dfrac{r\lambda}{\nu}\kappa'(\pi_e) \\
J_{3, 1}(\omega,  d)			& 0							& J_{3,3}(\omega, d)
\end{array}
\right]
\end{equation}
with 
\begin{align}
J_{1,1}(\omega,d)&=\Phi(\lambda)-\alpha-(1-\gamma)i(\omega,y_d) -(1-\gamma)\omega\left[\eta_p m+\eta_q\left(c_1-\frac{\kappa'(\pi_e)}{u_0}\right)\right] \\
J_{1,3}(\omega,d)&=-(1-\gamma)\omega\eta_q\left(c_2-r\frac{\kappa'(\pi_e)}{u_0}\right) \\
\label{eq:J_dom}
J_{3,1} (\omega,  d)
&= d\left[ \frac{\kappa'(\pi_e)}{\nu}-\eta_p m-\eta_q\left(c_1-\frac{\kappa'(\pi_e)}{u_0}\right)\right]+(1-c_1) \\
\label{eq:J_dom2}
J_{3,3} (\omega, d)
&= \left[r-c_2- \frac{\kappa(\pi_e)}{\nu}+\delta -i(\omega,y_d) \right] + d\left[\frac{\kappa'(\pi_e)}{\nu}-\eta_q\left(c_2-r\frac{\kappa'(\pi_e)}{u_0}\right)\right].
\end{align}

Under technical conditions similar to those found in \cite{GrasselliCostaLima2012, CostaLimaGrasselliWangWu2014, GrasselliNguyenHuu2015}
we shall establish that the model \eqref{keen_demand_inflation} exhibits three meaningful types of equilibrium points, analogous to those 
described in \cite{GrasselliNguyenHuu2015}.
The first one is a good equilibrium, corresponding to a desirable situation with finite debt and positive wages and employment.
The second one is a debt crisis equilibrium corresponding to a finite level of debt and vanishing wages and employment. The third one is a deflationary state that can either accept credit explosion or not, which appears with the introduction of price dynamics.
A fourth situation corresponding to the trivial equilibrium $(\omega,\lambda,d)=(0,0,0)$ is theoretically possible, but always unstable and therefore irrelevant in the present context. The following results rely on a standard
equilibrium analysis with Hartman-Grobman theorem and  can be skipped at first reading.

We start by assuming that $\kappa'(\pi_e)>0\;$ for all $\pi_e \in \R$ and satisfies  
\begin{equation}
\label{eq:kappa increasing}
\lim_{\pi_e \to - \infty}\kappa(\pi_p)=\kappa_0< \nu (\alpha+\beta + \delta)<\lim_{\pi_e \to + \infty} \kappa(\pi_e).
\end{equation}
We assume further that 
\begin{equation}
\label{eq:Phi in zero} 
\Phi'(\lambda)>0 \ \text{on}\ (0,1), \quad \Phi(0)< \alpha.
\end{equation}

\paragraph{Trivial equilibrium} 

At the equilibrium point $(\omega,\lambda,d)=(0,0,0)$, the Jacobian \eqref{eq:jacobian_keen_modified} becomes
\begin{equation*}
\label{eq:jacobian at zero}
\left[
\begin{array}{ccc}
J_{1, 1}(0, 0) 	& 0				& 0 \\
0		& \frac{\kappa(1)}{\nu}-\alpha-\beta-\delta	& 0 \\
1-c_1   & 0		& J_{3,3}(0,0) 
\end{array}
\right]
\end{equation*}
We therefore see that this point is unstable provided \eqref{growth_1} holds, which is 
likely to be true in practice, as already observed in Section \ref{non_monetary_long}.

\paragraph{Steady growth equilibrium}
\label{sec:good equilibrium 1}
Following  \cite{GrasselliCostaLima2012}, we call the equilibrium with finite debt
and positive wages and employment rate the \textit{good equilibrium} for \eqref{keen_demand_inflation}. We can see 
from the second equation in \eqref{keen_demand_inflation} that this is characterized by  
\begin{equation}
\kappa(1-\overline\omega_1 - r\overline d_1) = \nu(\alpha+\beta+\delta),
\end{equation}
which can be uniquely solved for $\overline\pi_1= 1-\overline\omega_1 - r\overline d_1$ because of condition \eqref{eq:kappa increasing}.
This corresponds to a steady state where the growth rate equals $ \alpha+ \beta$, the natural growth rate
of the economy.
Accordingly, $\overline d_1$ is a root of the following quadratic equation:

\begin{equation}
\label{eq:quadratic polynomial of d_p}
A_1d^2+A_2d+A_3=0
\end{equation}
where
\[A_1 =\eta_prm, \quad A_2 =\eta_p(1-m(1-\overline\pi_1))-\eta_q\nu(\alpha+\beta+\delta)+(\eta_q-1)(c_1r-c_2), \quad
A_3 =(1-\overline\pi_1)(1-c_1(\eta_q+1)).\]
Provided $\Delta:=A_2^2 - 4 A_1 A_3\ge 0$,
there is at least one real solution to \eqref{eq:quadratic polynomial of d_p} given by
\begin{equation}
\label{eq:d_pq}
{\overline d}^{\pm}_1 = \frac{-A_2 \pm \sqrt{\Delta}}{2\eta_prm}\;.
\end{equation}
It is rather difficult to get information out of this expression, and one shall
study the above value numerically.

We shall impose $\overline d_1<(1-\overline \pi_1 )/r$, in order to ensure that $\overline\omega_1>0$. Given $\overline\omega_1$, we obtain  
\begin{equation}
\overline\lambda_1 = \Phi^{-1} \left(\alpha + (1-\gamma)i(\overline\omega_1,\overline d_1)\right),
\end{equation}
which, on account of  \eqref{eq:Phi in zero}, always exists provided $i(\overline\omega_1,\overline d_1)>0$. At the point $(\overline\omega_1, \overline\lambda_1,  \overline d_1)$, the Jacobian \eqref{eq:jacobian_keen_modified} becomes
\begin{equation}
\label{eq:jacobian_keen_modified at equilibrium}
\left[
\begin{array}{ccc}
J_{1,1}(\overline\omega_1,\overline d_1)	& \overline\omega_1\Phi'(\overline\lambda_1) & 0 \\
-\dfrac{\overline\lambda_1}{\nu}\kappa'(\overline\pi_1)	& 0			& -\dfrac{r\overline\lambda_1}{\nu}\kappa'(\overline\pi_1) \\
J_{3, 1}(\overline\omega_1,\overline d_1)& 0	& J_{3,3}(\overline\omega_1,\overline d_1)
\end{array}
\right]\;.
\end{equation}
Computing the characteristic polynomial $P[X]$ for the matrix \eqref{eq:jacobian_keen_modified at equilibrium}
and applying the Routh-Hurwitz criterion provides a necessary and sufficient condition
for all the roots of a cubic polynomial to have negative real part,
which in turn ensure local stability for this equilibrium point. They boil down to the following two fairly non-intuitive conditions, which need to be checked numerically:
\begin{align}
\label{eq:routh hurwitz condition 1}
&J_{3,3}(\overline\omega_1,\overline d_1)   <\min \left\{-J_{1,1}(\overline\omega_1,\overline d_1) ; -\frac{\overline\lambda_1 
\kappa'(\overline\pi_1)\overline\omega_1 \Phi'(\overline\lambda_1)}{\nu J_{1,1}(\overline\omega_1,\overline d_1)} ; 
r J_{3,1}(\omega^1, d^1) \right\} \\
\label{eq:routh hurwitz condition 2}
&\frac{r J_{3,1}(\overline\omega_1,\overline d_1)-J_{3,3}(\overline\omega_1,\overline d_1)}
{J_{1,1}(\overline\omega_1,\overline d_1) +J_{3,3}(\overline\omega_1,\overline d_1)}
+\frac{\nu J_{3,3}(\overline\omega_1,\overline d_1)J_{1,1}(\overline\omega_1,\overline d_1)}
{\overline\lambda_1 \kappa'(\overline \pi_e)\overline\omega_1 \Phi'(\overline\lambda^1)}
>-1.
\end{align}


At equilibrium, we find that demand equals 
\begin{equation}
\overline y_d = c_1\overline\omega_1+c_2\overline d_1+\frac{\nu(\alpha+\beta+\delta)}{u_0}
\end{equation}
When  $\overline y_d>1$, this equilibrium is not economically meaningful, 
since in this case \eqref{v_demand} leads to vanishing inventories 
in finite time and the model ceases to make sense. Accordingly, this restricts the constant 
capacity utilization $u_0$ to the range 
\begin{equation}
u_0\geq \frac{\nu(\alpha+\beta+\delta)}{1-c_1\overline\omega_1-c_2\overline d_1},
\end{equation}
in which case the relative inventory level $\v=V/Y$ converges to the equilibrium value
\begin{equation}
\label{inventory_ratio}
\overline\v_1=\frac{1-c_1\overline\omega_1-c_2\overline d_1-\nu(\alpha+\beta+\delta)/u_0}{\alpha+\beta},
\end{equation}
with the Keen model of \cite{Keen1995} corresponding to structurally unstable 
special case with $\overline y_d=1$ and $\overline \v_1=0$.

\paragraph{Debt crisis}
\label{sec: bad equilibrium 1}

As described in \cite{GrasselliCostaLima2012}, a key feature of the original Keen model \cite{Keen1995} is 
that it admits an equilibrium of the form $(\omega,\lambda,d)=(0,0,+\infty)$, that is to say, corresponding 
to unbounded growth in the debt ratio at the same time that the wage share and employment rate decrease 
to zero. The analysis in \cite{GrasselliCostaLima2012} was done using a change the variable $q=1/d$ and 
observing that $(\omega,\lambda,q)=(0,0,0)$ becomes an equilibrium point for the transformed system. 

In the present case, we see that this change of variable turns the third equation in \eqref{keen_demand_inflation} into
\begin{equation*}
\dot{q} = q \left[\frac{\kappa(u_0,\pi_e)}{\nu}-\delta(u_0)+i(\omega,1/q)-r\right] - q^2 \left[(1-c_1)\omega-\frac{c_2}{q}\right].
\end{equation*}
Using the price dynamics \eqref{inflation}, with demand given by the \eqref{eq:demand share} and consumption of 
the form $\theta(\omega,d)=c_1\omega+c_2d$ we find that the term $q i(\omega,1/q)$ in the expression above becomes
\begin{equation*}
q i(\omega,1/q)=q\left[\eta_p(m\omega-1)+\eta_q\left(c_1\omega+\frac{\kappa(u_0,\pi_e)}{u_0}-1\right)\right]+\eta_qc_2,
\end{equation*}
from which we deduce that $q=0$ does not lead to $\dot{q}=0$ in the transformed system. Observe that a similar problem arises 
with the term $\omega i(\omega,1/q)$ originating from the first equation in \eqref{keen_demand_inflation} whenever $\gamma\neq 1$. More generally, we conclude that the system \eqref{keen_demand_inflation} does {\em not} exhibit an equilibrium characterized 
by $d\to+\infty$ provided the consumption function has a wealth effect that grows at least linearly in $d$. 

On the other hand, system \eqref{keen_demand_inflation} admits a different type of debt crisis corresponding to no economic activity and a finite debt ratio, 
namely, an equilibrium of the form $(\omega,\lambda,d)=(0,0,\overline d_2)$ where $\overline d_2$ satisfies 
\begin{equation}
\frac{\kappa(1-r\overline d_2)}{\nu}-\delta=r-c_2-i(0,\overline d_2)
\end{equation}
with
\begin{equation}
\label{inflation_2}
i(0,\overline d_2)=-\eta_p+\eta_q\left(c_2\overline d_2+\frac{\kappa(1-r\overline d_2)}{u_0}-1\right).
\end{equation}

The Jacobian matrix \eqref{eq:jacobian_keen_modified} at point
$(0,0, \overline d_2)$ becomes
\begin{equation*}
\label{eq:jacobian at trivial}
\left[
\begin{array}{ccc}
\Phi(0)-\alpha-(1-\gamma)i(0,\overline d_2)	& 0				& 0 \\
0		& r-c_2-i(0,\overline d_2)-\alpha-\beta	& 0 \\
J_{3, 1}(0, \overline d_2)			& 0							& J_{3, 3}(0, \overline d_2)
\end{array}
\right] \;.
\end{equation*}
The conditions for local stability of $(0,0, \overline d_2)$ then read
\begin{equation}
\Phi(0)-\alpha-(1-\gamma)i(0,\overline d_2)<0, \quad r-c_2-i(0,\overline d_2) < \alpha+\beta, \quad J_{3,3}(0,\overline d_2) < 0.
\end{equation}
Because \eqref{inflation_2} cannot be solved exactly, all three conditions must be checked numerically. That is to say, local stability of the debt crisis 
equilibrium $(0,0, \overline d_2)$ cannot be ruled out a priori.

\paragraph{Deflationary equilibria}
Another undesired type of equilibrium, similar to the new one found in \cite{GrasselliNguyenHuu2015}, appears in the monetary version of the long-run dynamics
and corresponds to a very specific situation when decreasing real wages, due to low employment rate, are compensated by deflation in the economy.
Similar to Section \ref{keen_inflation_section}, these equilibria are of the form $(\overline\omega_3,0, \overline d_3)$ where $\overline\omega_3$ and $\overline d_3$ are solutions to the nonlinear equations
\begin{align}
i(\overline\omega_3,\overline d_3)&=\frac{\Phi(0)-\alpha}{1-\gamma} \\
(c_1-1)\overline\omega_3&=\overline d_3\left[r-c_2-\frac{\kappa(1-\overline\omega_3-r\overline d_3)}{\nu}+\delta-i(\overline\omega_3,\overline d_3)\right]
\end{align}
where
\begin{equation}
i(\overline\omega_3,\overline d_3)=\eta_p(m\overline \omega_3-1)+\eta_q(c_1\overline\omega_3+c_2\overline d_3-1) +\eta_q\frac{\kappa(1-\overline\omega_3-r\overline d_3)}{u_0}.
\end{equation}
Observe that \eqref{eq:Phi in zero} implies that $i(\overline\omega_3,\overline d_3)<0$, confirming that this 
equilibrium corresponds to a deflationary economy. 

The Jacobian matrix \eqref{eq:jacobian_keen_modified} at this point, $J(\omega^3, 0,  d^3)$ becomes
\begin{equation}
\label{eq:jacobian_keen_modified new}
\left[
\begin{array}{ccc}
J_{1,1}(\overline\omega_3,\overline d_3)	& \omega^3 \Phi'(0)	& 0 \\
0	& \dfrac{\kappa(1-\overline\omega_3-r\overline d_3)}{\nu}-\alpha-\beta-\delta			& 0 \\
J_{3, 1}(\overline\omega_3,\overline d_3) & 0	& J_{3,3}(\overline\omega_3,\overline d_3)
\end{array}
\right].
\end{equation}
Inverting the order of $\omega$ and $\lambda$ thus provide a lower triangular matrix, from 
which we can immediately obtain the three eigenvalues. Assuming $\overline\omega_3>0$, the conditions for local stability of $(\overline\omega_3,0, \overline d_3)$
then read
\begin{equation}
\eta_p m+\eta_q\left(c_1-\frac{\kappa'(1-\overline\omega_3-r\overline d_3)}{u_0}\right)>0, \quad 
\frac{\kappa(1-\overline\omega_3-r\overline d_3)}{\nu}-\delta < \alpha+\beta, \quad
J_{3,3}(\overline\omega_3,\overline d_3) < 0,
\end{equation}
all of which must be checked numerically. In other words, local stability of the deflationary equilibrium $(\overline\omega_3,0, \overline d_3)$ cannot 
be ruled out a priori either.

\section{Short-run dynamics}
\label{sec:short-run}

\subsection{Supply-demand t\^atonnement}

In contrast with Section \ref{sec:effective demand in the keen model}, we now consider what happens when {\em only} short-run effects are taking into account. For this, 
suppose that $\alpha+\beta=0$, so that the long-run equilibrium corresponds to an economy with zero growth. In this case, expected sales and investment in 
planned inventory should be driven solely by short-run fluctuations in demand, and we accordingly set 
\begin{equation}
g_e(u,\pi_e)=0.
\end{equation}
Inserting this into  \eqref{inflation}, \eqref{eq:v}, and \eqref{output_dynamics_franke}, we see that the inventory ratio and instantaneous growth rate of the economy become 
\begin{equation}
\v = \frac{[1+f_d\eta_d]y_e - 1}{\eta_d}, \quad 
g(y_e, y_d) = \eta_e(1+f_d \eta_d) (y_d -y_e) + \eta_d(y_d - 1).
\end{equation}
Similarly, because the long-run equilibrium level for capital is constant, investment is only necessary in order to replace depreciated capital, so that we can 
set
\begin{equation}
\kappa(u,\pi_e)=\nu\delta(u).
\end{equation}
The system \eqref{franke} then becomes
\begin{equation}
\label{eq:fast system}
\left\{
\begin{array}{l}
\dot{\omega} = \omega[\Phi(\lambda)-(1-\gamma) i(\omega,y_d,y_e)] \\
\dot{\lambda} = \lambda g(y_e, y_d) \\
\dot{d} = d[r-g(y_e, y_d)-i(\omega,y_d,y_e)] + \omega-\theta(\omega,d) \\
\dot{y}_e = -y_e g(y_e, y_d) + \eta_e (y_d - y_e) \\
\dot{u} = u g(y_e, y_d) 
\end{array}
\right.
\end{equation}

To proceed with the analysis further, we now consider the type of short business cycles first reported in \cite{Kitchin1923}, which are attributed to 
the response of firms to changes in inventory, but with lags in information. Accordingly, we focus exclusively on the fluctuations generated by 
the adjustment terms related to the parameters $\eta_e,\eta_d,\eta_q$ and set $\eta_p=0$ and $\Phi(\cdot)\equiv 0$, thereby ignoring both the labour cost  
push in \eqref{inflation} and the effect of employment in the wage bargaining equation \eqref{eq:wage 3}. This leads to 
\begin{equation}
i(\omega,y_d,y_e)=i(y_d,y_e)=\eta_q(y_d-y_e) .
\end{equation}
Assuming further that $\delta(u)=\delta u$ for $\delta>0$ and that 
\begin{equation}
\theta(\omega,d)=c_1\omega+c_2 d=c_1 \omega \qquad (\text{i.e. } c_2=0)
\end{equation}
we find that $y_d=c_1\omega+\nu\delta$, so that the growth rate of $y_d$ is the same as the one of $\omega$. We then obtain that 
\eqref{eq:fast system} decouples into a two-dimensional system for the variables $y_d$ and $y_e$ of the form 
\begin{equation}
\label{short-run}
\left\{
\begin{array}{ll}
\dot{y}_d &= - (1-\gamma)y_d  \eta_q(y_d-y_e) \\
\dot{y}_e	&= \eta_e (y_d - y_e) - y_e g(y_e, y_d)
\end{array}
\right.
\end{equation}
and an auxiliary system for $(\omega,\lambda,d)$ that can be solved after $(y_d,y_e)$ have been determined. 

This is a situation with constant level of capital,
and investment answering to running costs and depreciation only.
The productivity and the population size are assumed to be constant.
Firms adjust their expectations by replicating the demand with relaxation parameter $\eta_e$.
Inventories aim at being a fraction $f_d$ of that expected demand, with relaxation parameter $\eta_d$.
Prices move only to accommodate inventories, via the parameter $\eta_q$.
Wages respond to inflation though the parameter $\gamma$ and the aggregate real purchasing power is affected by $i(y_d,y_e)$.
We are thus in a strictly short-run supply-demand t\^atonement mechanism.

\subsection{Phase plan analysis}
\label{ap:inventory analysis}

It is easy to see that $\{y_d\ge0\}$ is invariant by \eqref{short-run},
which induces that $\{y_e\ge 0\}$ is also invariant.
Nothing, however, bounds either these two variables or the auxiliary function $g(y_e, y_d)$ from above.
Consequently, great care shall be taken in the interpretation of the model dynamics,
since it allows for $\lambda>1$, $u>1$ or $\v<0$ without any direct feedback.

We rewrite the system \eqref{short-run} into the following form
\begin{equation}
\label{short-run new form}
\left\{
\begin{array}{ll}
\dot{y}_d &= -(1-\gamma) y_d  \eta_{q}(y_d-y_e) \\
\dot{y}_e	&= y_e \eta_0 (y_e - y_e^-) - y_d (\eta_d + \eta_0)(y_e - y_e^+)
\end{array}
\right.
\end{equation}
where $\eta_0 = \eta_e (1+f_d \eta_d)$ and
$$
y_e^- = \frac{\eta_e - \eta_d}{\eta_0}<y_e^+ = \frac{\eta_e}{\eta_d + \eta_0}< 1.
$$
The above inequalities hold for any positive parameters configuration.
The isocline $\dot y_d=0$ is given by the lines $y_d=0$ and $y_e=y_d$, and it easily follows 
that $\dot{y}_d<0$ on $\{0\le y_e<y_d\}$ and $\dot{y}_d>0$ on $\{y_e>y_d\}$.

On the other hand, the isocline $\dot y_e=0$ is given by the following two functions of $y_d$:
$$
n_{\pm}(y_d) = \frac{(\eta_d+\eta_0)y_d + \eta_0 y_e^-}{2\eta_0}\pm \frac{1}{2\eta_0}\sqrt{\Delta(y_d)},
$$
with 
$
\Delta(y_d)=\left((\eta_d+\eta_0)y_d + \eta_0 y_e^-\right)^2 - 4 \eta_0 y_e^+ (\eta_d+\eta_0)y_d.
$
The polynomial $\Delta(y_d)$ itself has two real roots $y_d^-<y_d^+$ given by
$$
y_d^\pm = \frac{\eta_0}{\eta_0+\eta_d}\left(2y_e^+ - y_e^- \pm2 \sqrt{y_e^+ (y_e^+-y_e^-)}\right).
$$
It is easy to see that $y_d^->0$, since $\Delta(0) = (\eta_0 y_e^-)^2>0$, and that $y_d^+<1$, since $\Delta(1) = (\eta_d f_d \eta_e)^2>0$. Moreover, 
on the interval $(y_d^-, y_d^+) \subset (0,1)$ we have that $\Delta(y_d)<0$, so that the curves $n_{\pm}(y_d)$ are not defined and $\dot{y}_e>0$.  
Alternatively, on $[0, y_d^-)\cup(y_d^+, +\infty)$, we have that $\dot{y}_e<0$ for $y_e \in (n_-(y_d), n_+(y_d))$ and
$\dot{y_e}>0$ for $y_e \in (0,n_-(y_d))\cup (n_+(y_d), +\infty)$.

\begin{figure*}
	\centering
	\includegraphics[width = 0.95\textwidth]{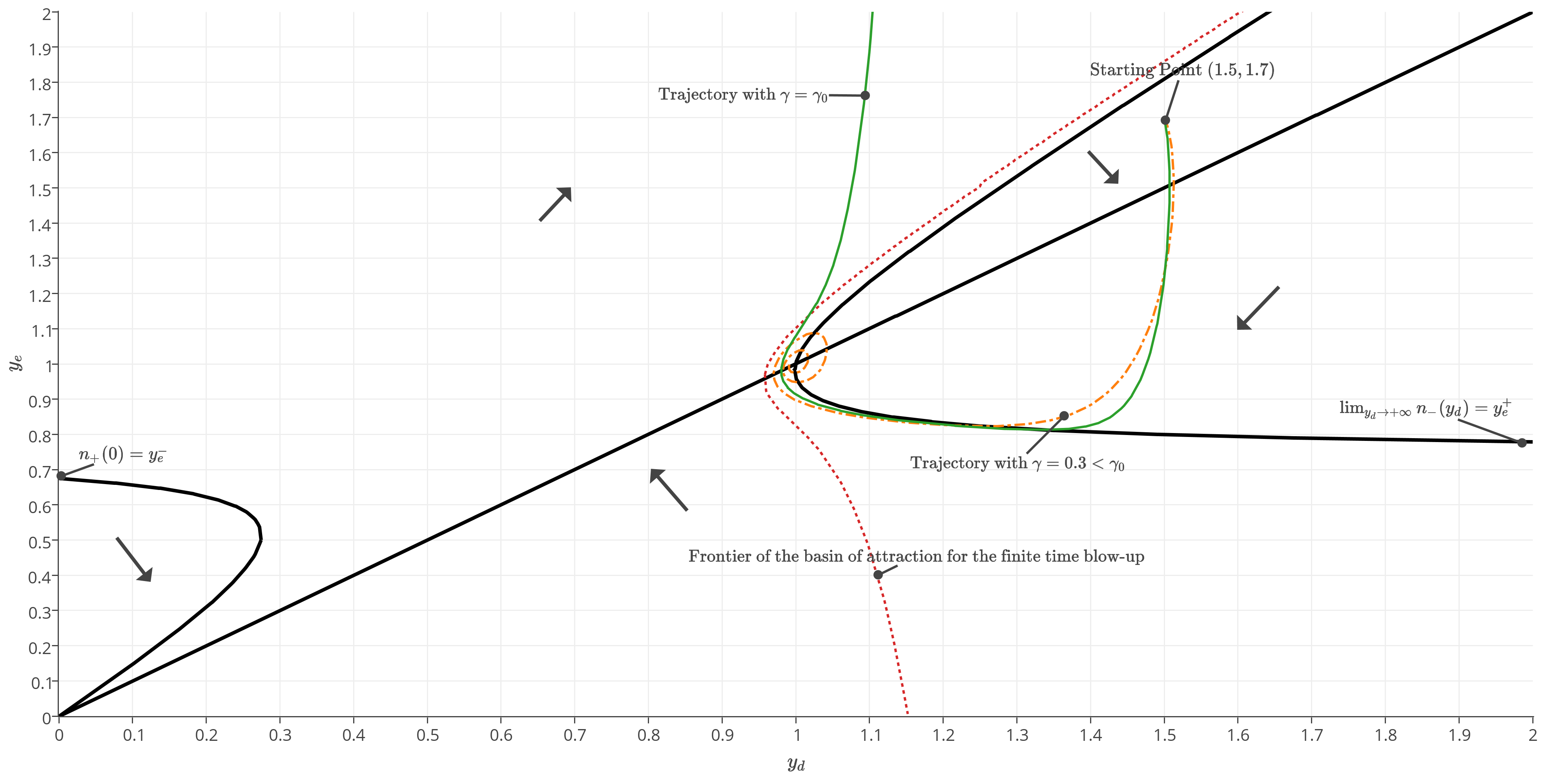}
	\caption{Discriminant regions for directional quadrants of the vector field \eqref{short-run} [black lines, black arrows].
		Sample phase trajectories given with initial point $(y_d, y_e)=(1.5, 1.7)$ for $\gamma=0.3$ [orange dotted-line] and $\gamma=\gamma_0 = 0.625$ [green line].
		Numerical approximation of the basin of attraction of the finite-time blow-up [dotted red line] with $\gamma=0.3$.
		Parameters $(\eta_e, \eta_d, \eta_q, f_d)=(2.5, 0.75, 0.25, 0.05)$.}
	\label{fig:phase}
\end{figure*}

Notice that $y_e = y_e^-$ or $y_e=y_e^+$ imply $\dot{y_e}>0$ for all $y_d$.
This actually provides limit cases $n_+(0)=y_e^-$ and $\lim_{y_d \to +\infty} n_-(y_d)=y_e^+$.
In addition,
we can show that 
$$
 \frac{\partial n_-}{\partial y_d}(0) = \frac{\eta_0 + \eta_d}{\eta_0}\left(\frac{3}{2}+2\frac{\eta_d(\eta_d + \eta_0 - \eta_e)}{(\eta_d + \eta_0)(\eta_e - \eta_d)}\right)
$$
which is greater than one provided $\eta_e>\eta_d$. 
In that case, we find that $\{\dot{y}_e<0, y_d \le y_d^-\}\cap \{y_d = y_e\} = \emptyset$. On the other hand, for large values of $y_d$, we have that
\begin{equation}
\label{super}
n_+(y_d) = \frac{\eta_d + \eta_0}{\eta_0}y_d + \mathrm{o}(|y_d|) \text{ for } y_d \to +\infty,
\end{equation}
so that $\{ \dot{y}_e>0, y_d> y_d^+ \}\cap \{ y_d = y_e\} = \emptyset$.
This is summed up by black thick lines and arrows giving the directional quadrants of the system in each area on Figure \ref{fig:phase}.

\subsection{Equilibrium points}
\label{sec:equilibrium points 2D}

Three equilibria are possible for \eqref{short-run new form}. 
The first equilibrium $(\overline{y}_d, \overline{y}_e)=(1,1)$ corresponds to the interior equilibrium \eqref{y_bar_franke} in the general model with 
$(\alpha+\beta)=0$. At this equilibrium, expected sales equals output and equals demand: the t\^atonnement is successful.
Moreover, the equilibrium inflation rate and the growth rate are both zero, $\lambda$ and $u$ are constant, and inventory provides
a constant buffer $\v=f_d$ between production and sales. 

The Jacobian matrix of \eqref{short-run new form} is
$$
\begin{bmatrix}
-(1-\gamma)\eta_q (2y_d - y_e) & (1-\gamma)\eta_q y_d\\
\eta_e - y_e (\eta_0 + \eta_d) & \eta_d - \eta_e + 2\eta_0 y_e - (\eta_0+\eta_d)y_d
\end{bmatrix}.
$$
At the first equilibrium, it simplifies into
$$
\begin{bmatrix}
-(1-\gamma)\eta_q & (1-\gamma)\eta_q \\
\eta_e - (\eta_0 + \eta_d) & \eta_0 - \eta_e
\end{bmatrix}
$$
which yields the characteristic equation:
$$
X^2 + X \left[(1-\gamma)\eta_q - \eta_e f_d \eta_d\right] + (1-\gamma)\eta_q \eta_d = 0.
$$
All parameters being positive, and $\gamma< 1$,
the real part of the two roots is negative if and only if (Routh-Hurwitz criterion)
\begin{equation}
\label{condition negative real part}
\gamma< \gamma_0:= 1-\eta_e \eta_d f_d/\eta_q.
\end{equation}
This equilibrium is thus locally attractive if \eqref{condition negative real part} holds.
When $\gamma=\gamma_0$,
the roots of the characteristic equations are purely imaginary and the system
undergoes an Andronov-Hopf bifurcation.
The first Lyapunov exponent at the bifurcation value $\ell_1(\gamma_0)$
is positive, 
so that the bifurcation is sub-critical: the limit cycle is unstable (this is confirmed by simulations).
When $\gamma\ge \gamma_0$,
in full generality, the equilibrium point is locally repelling.

The second equilibrium is given by $(\overline{y}_d, \overline{y}_e)=(0,0)$
and means the collapse of the market, with all variables dwindling to zero. 
At this point the Jacobian matrix is
$$
J(0,0) = 
\begin{bmatrix}
0 & 0 \\
\eta_e & \eta_d - \eta_e
\end{bmatrix},
$$
and we see that this equilibrium is unstable for $\eta_d>\eta_e$ and fails to be asymptotically stable, even if $\eta_d<\eta_e$, since
the kernel of the Jacobian is of dimension one.

If the system does not converge toward $(1,1)$ nor $(0,0)$,
there exists a singular attractive state for \eqref{short-run} characterized by a finite-time blow-up towards $(+\infty,+\infty)$. 
By writing the system under the form $(h,x) = (1/y_d, y_d/y_e)$,
we obtain the dynamics
\begin{equation}
\label{short-run inverse}
\left\{
\begin{array}{ll}
\dot{h} =& (1-\gamma)\eta_q\left(1- \frac{1}{x}\right)  \\
\dot{x} =& \left((1-\gamma)\eta_q + \eta_0\right)\frac{-1}{h} + (\eta_0+\eta_d + (1-\gamma)\eta_q)\frac{x}{h} - (\eta_0 - \eta_d)x - \eta_e x^2 .
\end{array}
\right.
\end{equation}
We are interested in the finite-time blow-up for both $h$ and $x$ toward $0$,
since $n_+(y_d)$ is super-linearly increasing in $y_d$, according to \eqref{super}.
We are able to express \eqref{short-run inverse} as a second order non-linear differential equation:
\begin{equation}
\ddot{h} = (1-\gamma)\eta_q \eta_d \left(\frac{1}{h}-1\right) - (\eta_e - \eta_d)\dot{h} + \frac{\dot{h}}{h}\left(\eta_0 -\eta_d - (1-\gamma)\eta_q\right)
 \quad + \frac{\dot{h}^2}{h}\left(1 - \frac{\eta_0}{(1-\gamma)\eta_q}\right). \label{ode h}
\end{equation}
and the local behavior defined by $\lim_{h,x\to 0}\dot{h} = -\infty$.
In that case, factorizing by $\dot{h}^2 /h$ allows to approximate \eqref{ode h} around that point by
$$
\ddot{h} \sim_{h\downarrow 0}  \frac{\dot{h}^2}{h}\left(1 - \frac{\eta_0}{(1-\gamma)\eta_q}\right).
$$
This is a generalized Emdem-Fowler equation \cite{polyanin1995handbook},
whose solution is of the form 
$$h(t) = a_2 \left(\frac{\eta_0}{(1-\gamma)\eta_q} t + a_1\right)^{\frac{(1-\gamma)\eta_q}{\eta_0}},$$ 
reaching zero in finite time for any parameters $a_1, a_2>0$.
This situation translates into a singular behaviour:
when expectations and demand are very high,
inventories must decrease to answer expected sales,
and prices are pushed down to control unplanned inventories.
However this stimulates the aggregate demand,
which in turns boost the expectations.
The growth rate of $Y$ shrinks and the situation gets worse.

\subsection{Price adjustment and information lag: another view}
\label{sec:another view}

In the present case,
assumption \eqref{inflation} takes on a major role about firms speed of price adjustment,
where it is assumed that they take into account direct unplanned inventory investment.
This has a significant impact on the stability of a supply-demand equilibrium.

\begin{figure*}
	\centering
	\includegraphics[width = 0.95\textwidth]{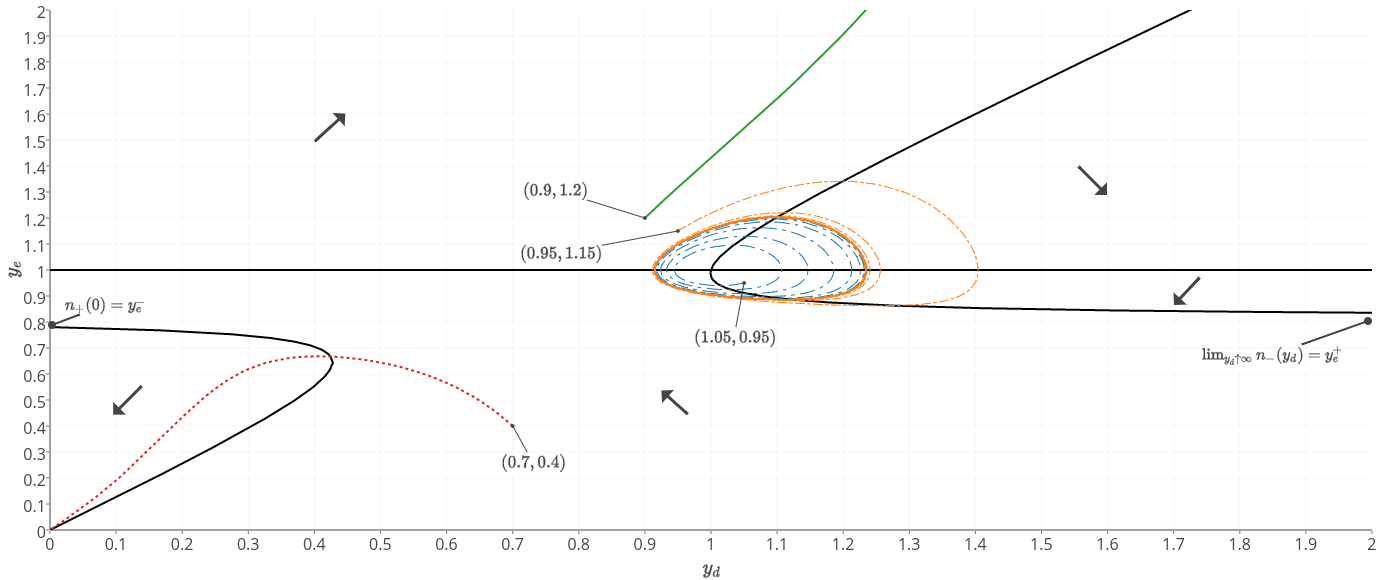}
	\caption{Discriminant regions for directional quadrants of the vector field \eqref{short-run alternative} [black lines, black arrows].
		Several sample phase trajectories given with initial points $(y_d, y_e)$.
		Parameters $(\eta_e, \eta_d, \eta_q, f_d, \gamma)=(2.5, 0.5, 0.25, 0.05, 0)$.}
	\label{fig:phase2}
\end{figure*}

Consider an alternative assumption to \eqref{inflation} with prices adjusting
with the difference between desired and observed inventory levels:
\begin{equation}
\frac{\dot{p}}{p} = \eta_p \left(m \frac{c}{p} - 1\right) - \eta_{q}\frac{V_d-V}{Y}  \label{eq:price-5}
\end{equation}
which, along with previous assumptions, provides the inflation rate $i(y_e)=\eta_q (1-y_e)/ \eta_v$.
Computations provide an alternative short-run business cycles system of the form
\begin{equation}
\label{short-run alternative}
\left\{
\begin{array}{ll}
\dot{y}_d &= - (1-\gamma)y_d  \frac{\eta_q}{\eta_v}(1-y_e) \\
\dot{y}_e	&= \eta_e (y_d - y_e) - y_e g(y_e, y_d),
\end{array}
\right.
\end{equation}
that we compare with \eqref{short-run}.
The slight difference with the latter concerns the first equation,
for which the isocline is given by $\{y_e=1\}$ instead of $\{y_d=y_e\}$.
Equilibrium points are interestingly the same, i.e.,
a supply-demand equilibrium $(1,1)$,
a market collapse $(0,0)$,
and a finite-time blow-up toward $(+\infty,+\infty)$.
The impact on stability is however much more involved.
The Jacobian for the first equilibrium is given by
\begin{equation}
\label{eq:simplified jacobian eq 1}
\begin{bmatrix}
0 & (1-\gamma) \eta_q/\eta_d\\
\eta_e - \eta_0 - \eta_d & \eta_0-\eta_e
\end{bmatrix}
\end{equation}
Since all entries of the matrix above are positive, both root of the characteristic equation are positive.
This equilibrium point is thus locally repelling. We observe on Figure \ref{fig:phase2} that points starting in the close neighborhood of $(1,1)$
converge to a semi-stable limit cycle.
Trajectories starting in the external neighborhood of that limit cycle
converge also numerically to that limit cycle, so that
the latter is locally stable.
These assertions are supported by the two trajectories starting
respectively at $(1.05, 0.95)$ and $(0.95, 1.15)$ on Figure \ref{fig:phase2}.
This seems to correspond strongly to a Kitchin cycle \cite{Kitchin1923},
where lag in information and decision adjustment (through parameters $\eta_d$, $\eta_e$ and $\eta_q$)
affect prices, output, demand, inventory and employment in a periodic manner.
This cycle is locally stable, that is, it attracts a whole region including it.

At the bad equilibrium $(0,0)$ the Jacobian equals
\begin{equation}
\label{eq:simplified jacobian eq 0}
J(0,0) = 
\begin{bmatrix}
-(1-\gamma)\eta_q/ \eta_d & 0\\
\eta_e & \eta_d - \eta_e
\end{bmatrix}
\end{equation}
where the two eigenvalues are negative if and only if $\eta_d <\eta_e$.
Therefore, this point can be locally attractive, as the trajectory starting at $(y_d, y_e)=(0.7, 0.4)$
shows on Figure \ref{fig:phase2}. Here, the low level of inventories pushes prices up,
and pulls demand down, in a spiral dynamics toward a complete shut-down of economical activities:
employment and utilization dwindle to zero, and the market shuts down. 

The finite-time blow-up is similar to the previous case. See for example the trajectory starting at $(y_d, y_e)=(0.9, 1.2)$ on Figure \ref{fig:phase2}.

\section{Conclusions}
\label{sec:conclusion}

In the present article, we have presented a general, albeit complex, stock-flow consistent model for inventory dynamics in a closed monetary economy.  
The model relies heavily on adapted behaviour of firms regarding expected sales and desired inventory levels. To gain insight, we analyze the model in two specific limiting versions: a long-run dynamics ignoring the effect of instantaneous fluctuations in 
demand and a short-run one solely driven by these fluctuations.  

The long-run dynamics gives rise to a version of the Keen model \cite{Keen1995} where demand is not necessarily equal to output. 
This sheds light on the question of whether the rich set of trajectories obtained in \cite{Keen1995} and related models of debt-financed investment were 
an artefact of a strictly supply-driven model with no role for Keynesian effect demand. As the above analysis shows, one can relax the constraints 
of the Keen model by allowing an independently specified consumption function and still obtain broadly the same conclusions. The main 
difference is that the debt crisis, previously characterized by an explosive debt ratio, gets replaced by an equally bad equilibrium with a 
finite debt ratio but collapsing economy with vanishing wage share and employment rates. In both cases, Minskyan instability arising from financial charges
lead to the collapse of profits and an induced debt crisis.

The short-run dynamics reveals inventory cycles related to nominal rigidity of demand. For given speeds of expectations adjustment and inventory stocks,
a high degree of nominal illusion is necessary to ensure local stability. Yet, this situation can show divergence as the model does not include 
long-run feedback channels. A second dimension, illustrated by the difference in behavior in Sections \ref{sec:equilibrium points 2D} and \ref{sec:another view},
is the type of information on inventory used for price adjustment. By replacing unplanned inventory investment with 
mismatch in desired inventory levels as the factor determining price adjustments, we create a lag in information that gives rise
a stable limit cycle which we interpret as a Kitchin cycle. Nevertheless, this does not get rid of the divergent path, still possible if the situation is too far from the interior equilibrium. In addition, we have the possibility of a market failure where demand and supply meet at zero, for a condition that appears repeatedly in this section provided $\eta_d<\eta_e$. In other words inventories must adjust faster than expectations to avoid this type of crash.

In its closing sentences, \cite{Wen2011} asserts that ``general-equilibrium analysis of the business cycle with inventories is still in its infant stage." This is also true 
for disequilibrium analysis of the business cycle in the burgeoning recent literature on stock-flow consistent models \cite{CaverzasiGodin2013}, where inventories have received comparatively less attention than their more glamorous financial counterparts - cash balances, government deficits and the like. We hope this paper will help bring the analysis of inventory dynamics to a more diverse adolescence.  

\section*{Acknowledgements}
We are grateful to the participants of the Applied Stock-Flow Consistent Macro-modelling Summer School (Kingston University, August 1-8, 2016) where this work was presented. Partial financial support for this work was provided to MRG by the Institute for New Economic Thinking (Grant INO13-00011) and the Natural Sciences and Engineering Research Council of Canada (Discovery Grants) and to ANH by the Chair \textit{Energy and Prosperity, Energy Transition Financing and Evaluations}. 


\bibliographystyle{siam}
\bibliography{finance}

\end{document}